%% file: edge_based_stochastic_block_model_statistical_inference.tex
\documentclass{svproc}
\usepackage{graphicx}
\graphicspath{{Images/}}
\usepackage{color, soul}
\usepackage{amsmath, amsfonts, amssymb}
\usepackage{url}

\begin{document}
\mainmatter              
\title{Edge based stochastic block model statistical inference}
\titlerunning{}  
%
\author{Louis Duvivier\inst{1} \and R\'emy Cazabet\inst{2} \and C\'eline Robardet\inst{1}}
\authorrunning{Louis Duvivier et al.} 
%
%
\institute{
    Univ Lyon, INSA Lyon, CNRS, LIRIS UMR5205, F-69621 France \\
    \email{louis.duvivier@insa-lyon.fr}, \email{celine.robardet@insa-lyon.fr}, \\
    \and
    Univ Lyon, Universit\'e Lyon 1, CNRS, LIRIS UMR5205, F-69622 France \\
    \email{remy.cazabet@univ-lyon1.fr}
}

\maketitle

\begin{abstract}
Community detection in graphs often relies on ad hoc algorithms with no clear specification about the node partition they define as the best, which leads to uninterpretable communities. Stochastic block models (SBM) offer a framework to rigorously define communities, and to detect them using statistical inference method to distinguish structure from random fluctuations. In this paper, we introduce an alternative definition of SBM based on edge sampling. We derive from this definition a quality function to statistically infer the node partition used to generate a given graph. We then test it on synthetic graphs, and on the zachary karate club network.
\end{abstract}

\section{Introduction}
Since the introduction of modularity by Girvan and Newman \cite{girvan2002community}, it has been shown that many networks coming from scientific domain as diverse as sociology, biology and computer science exhibit a modular structure \cite{fortunato2016community}, in the sense that their nodes can be partitioned in groups characterized by their connectivity. Yet, there is no universal definition of a community. Many techniques and algorithms have been developed for detecting remarkable node partition in graphs, most of the time by optimizing a quality function which assigns a score to a node partition \cite{girvan2002community}, \cite{clauset2004finding}, \cite{newman2006finding}. The problem is that these algorithms rarely account for random fluctuations and it is thus impossible to say if the communities obtained reflect a real property of the graph under study or are just an artefact. In particular, it has been shown that even the very popular modularity may find communities in random graphs \cite{guimera2004modularity}.

Stochastic block models offer a theoretical framework to take into account random fluctuations while detecting communities \cite{peixoto2019bayesian}. Since they are probabilistic generative models, one can perform statistical inference in order to find the most probable model used to generate a given observed graph. The most common way to do this inference is to associate to each SBM the set of graphs it may generate: the larger the set, the smaller the probability to generate each of them \cite{peixoto_entropy_2012}, \cite{bianconi_entropy_2009}. This methodology based on the minimization of entropy has the strength of being rigorously mathematically grounded. Yet it suffers from one drawback: as it considers probability distributions on graph ensembles, the random variable considered is the whole graph. Thus statistical inference is performed on a single realization, which leads to overfitting. Although techniques have been introduced to mitigate this effect, it cannot be totally eliminated and it induces counter-intuitive behavior in some tricky situations \cite{duvivier2019minimum}.

In this paper, we propose a new quality function for node partitions, based on stochastic block models defined as probability distributions on a set of edges. This allows us to use statistical inference method in a more relevant way, relying on several realizations of the same random variable. To do so, we first define an edge-based stochastic block model, then use minimum description length method \cite{grunwald_tutorial_2004} to infer its parameters from an observed graph. Finally, we test this quality function on synthetic graphs, plus the Zachary Karate Club network.

\section{Methodology presentation}
\label{methodology presentation}

Traditionally, a stochastic block model is defined as a couple $(B, M)$, with $B$ a partition of the set of nodes $[1, n]$ in $p$ blocks $b_1, \dots, b_p$, and $M$ a $p \times p$ block adjacency matrix whose entries correspond to the number of edges between any two blocks (or equivalently to the density). These parameters define a set of generable graphs $\Omega_{B,M}$ from which graphs are sampled according to some probability distribution. As the probability distribution is defined on a set of graphs, we call the stochastic block models defined in this way generative models of graphs.

In this paper, we will consider stochastic block models as generative models of edges. It also takes as parameters a set of nodes $V = [1, n]$ partitioned in $p$ blocks $B = b_1, \dots, b_p$, but instead of a block adjacency matrix, it relies on a $p \times p$ block probability matrix $M$ such that:

\begin{itemize}
\item $\forall i, j, M[i,j] \in [0, 1] $
\item $\sum_{i,j} M[i,j] \times |b_i||b_j| = 1$
\end{itemize}

For a given partition $B$, the set of all matrices verifying those conditions will be denoted $\mathrm{Mat}(B)$. Given two nodes $u$ and $v$, belonging respectively to the block $b_i$ and $b_j$, the edge $u \rightarrow v$ is generated with probability $\mathbb{P}_{B, M}[u,v] = M[i,j]$. This probability distribution can be seen as a block-constant $n \times n$ matrix, and in the following, the notation $\mathbb{P}_{B,M}$ will refer indifferently to the probability distribution and to the corresponding matrix. We will also denote by $\mathrm{Prob\_mat(B)}$ the set of all $B$-constant edge probability matrices on $[1, n]^2$, defined as:

\begin{equation}
\mathrm{Prob\_mat}(B) = \{P \mid \exists M_P \in \mathrm{Mat}(B), P = \mathbb{P}_{B,M_P}\}
\end{equation}

Generating a graph $G = (V,E)$ made of $m$ edges $e_1, \dots, e_m$ with such a generative model of edges means generating each of its edges independently. Thus, $G$ is generated with probability:
\[
\mathbb{P}_{B,M}[G] = \prod_{i = 1}^m \mathbb{P}_{B,M}[e_i]
\]
In particular, this means that the same edge $u \rightarrow v$ can be sampled more than once, so for the rest of the paper we will work with multigraphs. To simplify computations, we will consider directed graphs with self-loops. In practice, we study a graph $G$ made of a set of vertices $V = [1, n]$ and a list $E$ of $m$ edges: $e_1 = u_1 \rightarrow v_1, \dots, e_m = u_m \rightarrow v_m$. We suppose that $G$ was generated by a stochastic block model $(B_0, M_0)$, thus that all edges in $E$ were independently sampled from the same probability distribution $\mathbb{P}_{B_0, M_0}$, and our objective is to identify the original parameters $B_0$ and $M_0$ used to generate $G$.

To do so, we rely on the minimum description length principle. This principle, borrowed from information theory, relies on the fact that any statistical regularity can be used for compression. Therefore, the quality of a statistical model can be measured by the compression it allows of the data under study. Let's give an example: Alice draws messages independently at random from a set $\Omega$, with a probability distribution $\mathbb{P}$ and she transmits them to Bob through a binary channel. Each message needs to be encoded through a coding pattern $C : \Omega \rightarrow \{0,1\}$. For any message $x \in \Omega$, we denote by $|C(x)|$ the length of its code. The expected length of the encoded message will then be:
\[ \underset{x \in \Omega}{\mathbb{E}}[|C(x)|] = \sum_{x \in \Omega} \mathbb{P}[x] \cdot |C(x)| \]
It can be shown that this expected value is minimum when $C$ is such that $\forall x, |C(x)| = -\mathrm{log}_2(\mathbb{P}[x])$, and in this case, the previous expression is called the entropy of $\mathbb{P}$. This result means that finding an optimal code $C^*$ and finding the original probability distribution $\mathbb{P}$ are the same problem, because: $\mathbb{P}[u,v] = 2^{-|C^*[u,v]|}$. This is what we will use to recover $\mathbb{P}_{B_0, M_0}$.

Let's suppose that Alice does not know $\mathbb{P}$, but that she can draw as many random messages as she wants from $\Omega$. Then, for any probability distribution $\mathbb{Q}$ on $\Omega$, she can define a code $C_Q$, under which the mean length of the messages $e_1, \dots, e_m$ transmitted will be:
\begin{equation}
  \mathrm{code\_len}(e_1, \dots, e_m, C_Q) = - \sum_{x \in \Omega} \frac{\#\{k \mid e_k = x\}}{m} \cdot \mathrm{log}_2(\mathbb{Q}[x])
\end{equation}
And, as we know that $\frac{\#\{k \mid e_k = x\}}{m} \underset{m \rightarrow \infty}{\longrightarrow} \mathbb{P}[x]$ because of the law of great numbers, it means that if $m$ is high enough, the best code $C^*$ will correspond to a distribution $\mathbb{Q}$ which will be a good approximation of $\mathbb{P}$.

In our case, the messages to be transmitted are the edges of $G$: $\{e_1, \dots, e_m\}$, drawn from the set $[1, n]^2$ with the probability distribution $\mathbb{P}_{B_0, M_0}$. We want to approximate this distribution, to deduce $B_0$ and $M_0$ from it, but to avoid overfitting, we do not minimize the encoding length of all edges $e_1, \dots, e_m$ at the same time, we consider them sequentially. It corresponds to a situation in which Alice observes the edges one at a time and transmits them right away, updating her code on the fly. At the other end, Bob updates his code in the same way. When Alice draws the $x^{th}$ edge, Bob only knows edges $e_1, \dots e_{x-1}$, so they optimize their code on this limited sample. For the remaining $m-x$ edges, as they have no information, they suppose they are random. Finally, as they know that edges are generated by a stochastic block model, they limit themselves to codes based on $B$-constant probability distributions, for some partition $B$. Thus $\mathbb{Q}^{B,x}$ is defined as:

\[\underset{\mathbb{Q} \in \mathrm{Prob\_mat}(B)}{\mathrm{argmin}} \left[\frac{x \cdot \mathrm{code\_len}(e_1,\dots, e_x, \mathbb{Q})}{m} - \sum_{u,v \in [1,n]} \frac{(m-x)}{m \cdot n^2} \cdot \mathrm{log}_2(\mathbb{Q}[u,v])\right]
\]
And the mean code length of the messages sent from Alice to Bob will be:

\begin{equation}
  \mathrm{code\_len}(E,B) = -\frac{1}{m} \sum_{x = 1}^m \mathrm{log}_2(\mathbb{Q}^{B, x-1}[e_x])
\end{equation}
Of course, it depends on the partition $B$ used by Alice and Bob. If we now imagine that each partition $B$ is tested in parallel, we can approximate $B_0$ by:

\begin{equation}
  B^* = \underset{B}{\mathrm{argmin}} \left(\mathrm{code\_len}(E,B)\right)
\end{equation}
This partition corresponds to the best sequential compression of edges $e_1, \dots, e_m$, and according to the minimum description length principle, it should correspond to the original partition $B_0$. It should be noted that sequential encoding suppose that edges are ordered, which is typically not the case (except for temporal graphs). Therefore, we need to choose an order, and it will necessarily be arbitrary. Yet, we observe in practice that, although it modifies the precise value of $\mathrm{code\_len}(E,B)$, fluctuations have a limited impact on the estimation $B^*$.

\section{Tests on synthetic graphs}

In order to test this estimator, we generated random graphs using edge-based stochastic block models, and observed how it behaves for various partitions of the nodes, in particular with underfitted and overfitted partitions. Before looking at the estimator itself, we investigated how the prediction probability of the next edge evolves as Alice draws more and more edges. Then, we tested how the mean code length behaves on partitions which are a coarsening or a refinement of the original partition, and on partitions with the same number of blocks as the original, but blocks of differents sizes, or shifted. Finally, we considered more complex SBM, with blocks of different sizes and density. 

\subsection{Prediction probability}

We start by considering three graphs $G_0$, $G_1$ and $G_2$. Each of them is made of $n = 128$ nodes and $m = 2800$ edges (density is about $0.17$), generated using three different stochastic block models described in table \ref{sbm}. 
\begin{table}
  \begin{center}
  \caption{\label{sbm}}
  \begin{tabular}{c|c|c}
    & Node partition & Block probability matrix \\
    \hline
    & & \\
    $S_0 = (B_0, M_0)$ & $([1, 128])$ & $\begin{bmatrix} \frac{1}{n^2} \end{bmatrix}$ \\[2ex]
    \hline
    & & \\
    $S_1 = (B_1, M_1)$ & $([1, 64], [65, 128])$ & $\frac{1}{n^2} \cdot \begin{bmatrix} 2 & 0 \\ 0 & 2 \end{bmatrix}$ \\[2ex]
    \hline
    & & \\
    $S_2 = (B_2, M_2)$ & $([1, 32], [33, 64], [65, 96], [97, 128])$ & $\frac{1}{n^2} \cdot \begin{bmatrix} 4 & 0 & 0 & 0 \\ 0 & 4 & 0 & 0 \\ 0 & 0 & 4 & 0 \\ 0 & 0 & 0 & 4 \end{bmatrix}$\\
  \end{tabular}
  \end{center}
\end{table}

\begin{figure}[t]
  \begin{center}
  \caption{Prediction probability against edge rank for three different graphs \label{prob_vs_x}}
  \includegraphics[width=\textwidth]{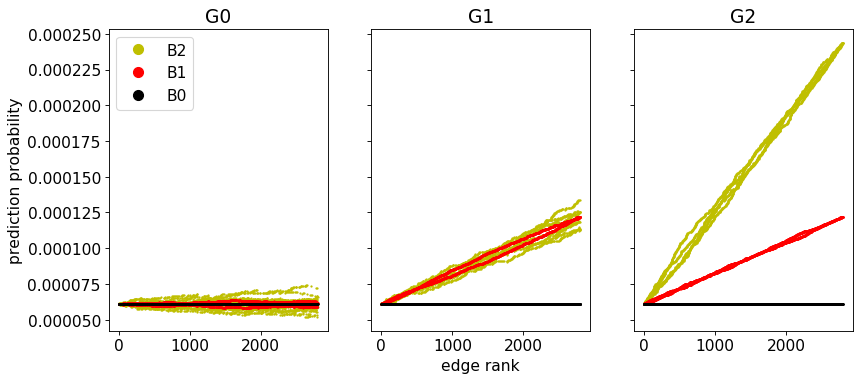}
  \end{center}
\end{figure}

For each of these graphs, we consider the prediction probability of the next edge $\mathbb{Q}^{B, x-1}[e_x]$ against $x$ for the three different partitions $B_0$, $B_1$ and $B_2$, which are all a refinement of the previous one. Results are shown on figure \ref{prob_vs_x}.

We observe for all three graphs, whatever $x$, the prediction probability based on the null partition $B_0$ is constant at $\frac{1}{128^2} \approx 0.00006$. This is logical, as the only $B_0$-constant probability matrix is the one corresponding to the uniform distribution. Therefore, $\forall x, \mathbb{Q}^{B_0, x} = \begin{bmatrix} \frac{1}{n^2} \end{bmatrix}$. For other partitions, the results depend on the graph. On $G_0$, generated with $B_0$ and thus presenting no block structure, the probability distributions associated to more refined partitions do no perform better than the one based on $B_0$. For some edges their prediction probability is better, but as often it is worse. On average, they have the same prediction power, they are only more sensible to the random fluctuations due to the order in which edges are drawn. On the other hand, for $G_1$, generated with $B_1$ (two blocks), we observe that refining the partition from one block to two allows the prediction probability to increase quickly. While it remains $\frac{1}{n^2}$ for the partition $B_0$, it rises up to $\frac{2}{n^2}$ for the partition $B_1$. Yet, refining even more the partition is worthless, as illustrated by the $B_2$ partition, with $4$ blocks, which does not bring any improvement on average. Finally, considering $G_2$, we observe that refining the partition brings more and more improvement to the prediction probability. With $B_0$ it remains stable at $\frac{1}{n^2}$, with $B_1$ it rises up to $\frac{2}{n^2}$, and with $B_2$ up to $\frac{4}{n^2}$.

\begin{figure}
  \begin{center}
    \caption{Mean prediction probability against partition refinement \label{mean_prob_vs_nb_com}}
    \includegraphics[width=0.8\textwidth]{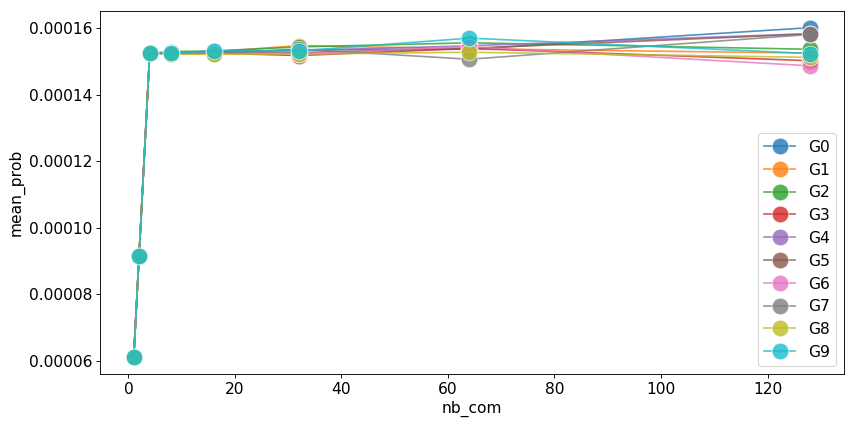}
    \caption{Mean code length against partition refinement: four communities graphs \label{mean_code_len_vs_nb_com}}
    \includegraphics[width=0.8\textwidth]{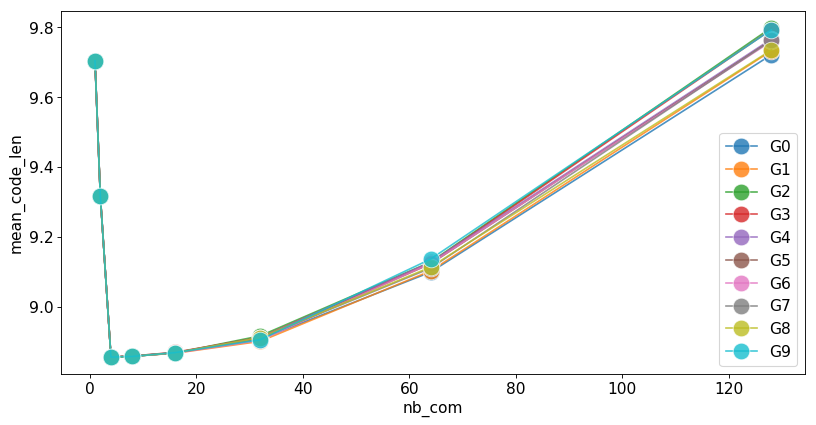}
    \caption{Mean code length against partition refinement: two communities graphs of various sharpness \label{mean_code_len_vs_nb_com_net}}
    \includegraphics[width=0.8\textwidth]{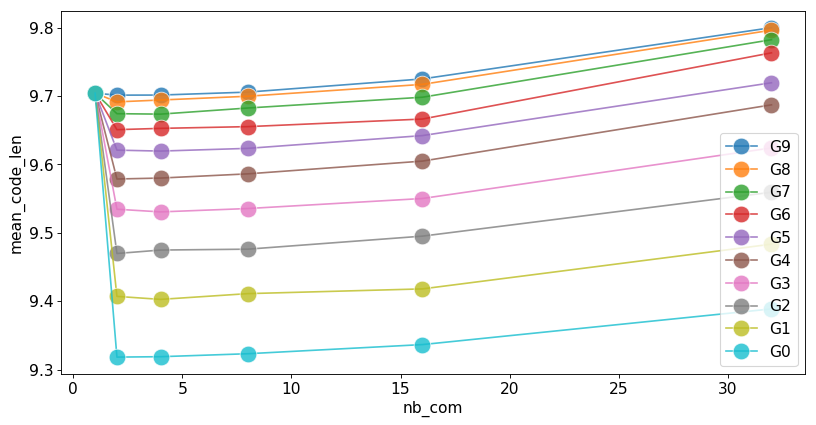}
  \end{center}
\end{figure}

To investigate further how the prediction probability evolves when refining the partition, we considered the mean prediction probability \[\mathrm{mean\_prob}(E, B) = \frac{1}{m} \cdot \sum_{x=1}^{m} \mathbb{Q}^{B, x-1}[e_x]\] on $10$ graphs generated with $S_2$. For each of them, the mean prediction probability is computed for $8$ different partitions $B_i$, with respectively $1$, $2$, $4$, $8$, $16$, $32$, $64$ and $128$ blocks. For each $i$, $B_{i+1}$ is obtained by dividing each block of $B_i$ in two blocks of equal size. The mean prediction probability is then plotted against the number of communities for each graphs on figure \ref{mean_prob_vs_nb_com}.

We observe that the mean prediction probability increases sharply as long as $B$ is coarser or equal to $B_2$ (the partition used to generate the edges). As soon as it becomes finer, it keeps increasing or decreasing a bit, according to the graph considered, but it mainly remains stable. This is due to the fact that $\mathbb{Q}^{B_i, x}$ tries to converge toward $M_2$. As $M_2$ does not belong to $\mathrm{Prob\_mat}(B_0)$ nor to $\mathrm{Prob\_mat}(B_1)$, its convergence is limited for these two partitions, and therefore the mean prediction probability is limited. On the other hand, for $i \geq 2$, $M_2 \in \mathrm{Prob\_mat}(B_i)$, so the convergence is not limited, but refining the partition is pointless, since $M_2$ is $B_2$-constant.

\subsection{Mean code length}

If we now plot the same curves, replacing the mean prediction probability by the mean code length, we obtain the results shown on figure \ref{mean_code_len_vs_nb_com}.

We observe that for all $10$ graphs, the mean code length sharply decreases until $i=2$. This is because, as we have just seen, for $i \leq 2$, refining the partition leads to a quick increase of $\mathbb{Q}^{B_i, x-1}[e_x]$, and therefore a decrease of the code length of the $x^{th}$ edges $-\mathrm{log}_2(\mathbb{Q}^{B_i, x-1}[e_x])$. On the other hand, for $i \geq 2$, the mean code length starts to increase again slowly and then faster, in contrast with the mean prediction probability that remained stable in this regime. We have seen that, as $i$ grows larger than $2$, $\mathbb{Q}^{B_i, x-1}[e_x]$ oscillates more and more due to random fluctuations. When computing the mean prediction probability, these oscillations compensate each other, but as logarithm is a concave function:
\[ - \mathrm{log}_2\left(\frac{1}{m} \cdot \sum_{x=1}^m \mathbb{Q}^{B_i, x-1}[e_x]\right) < -\frac{1}{m} \left(\sum_{x=1}^m \mathrm{log}_2(\mathbb{Q}^{B_i, x-1}[e_x]\right) \]
Therefore, the more $\mathbb{Q}^{B_i, x-1}[e_x]$ oscillates, the larger the mean code length in the end.

These two phenomenon are very important, because they explain how the mean code length as a quality function prevents both overfitting and underfitting. If the partition tested is too coarse with respect to the original partition, $\mathbb{Q}^{B,x}$ cannot converge toward the original block probability matrix, and the mean code length increases. On the other hand, if it is too fine, the convergence occurs but in a more noisy way, and this too leads to an increase of the mean code length.

Of course, it can work only if the edge generation probabilities are different enough and if the total number of edges drawn is large enough, for $\#\{k \mid u_k \in b_i \land v_k \in b_j\}$ to be significantly different from one pair of blocks $(b_i, b_j)$ to another. To illustrate this, we considered a set of $10$ graphs, still with $128$ nodes and $2800$ edges, generated by stochastic block models based on the partition $B_1$ (two blocks) and block probability matrices:
\[ M_i = \frac{1}{n^2} \cdot
\begin{bmatrix}
  (2 - \frac{i}{10}) & \frac{i}{10} \\
  \frac{i}{10} & (2 - \frac{i}{10})
\end{bmatrix}\]
Therefore, $S_0$ generates graphs with two perfectly separated communities, while $S_{9}$ generates graphs with almost no community structure. For each stochastic block model, we generate a graph $G_i$ and compute the mean code length for $6$ different partitions, $B_0$ to $B_5$, defined as before with $1$ to $32$ blocks. Results are plotted on figure \ref{mean_code_len_vs_nb_com_net}. We observe that for $i = 0, 2, 4, 6, 8$, the minimum mean code length is obtained for the two blocks partition $B_1$, while for the other, it is obtained for the four blocks partition $B_2$. This shows that fuzzy communities may lead to limited overfitting, but that the quality function is very robust against underfitting.

\begin{figure}
  \begin{center}
    \caption{Mean code length against cut (left) and offset (right) \label{mean_code_len_vs_cut_offset}}
    \includegraphics[height=0.28\textheight]{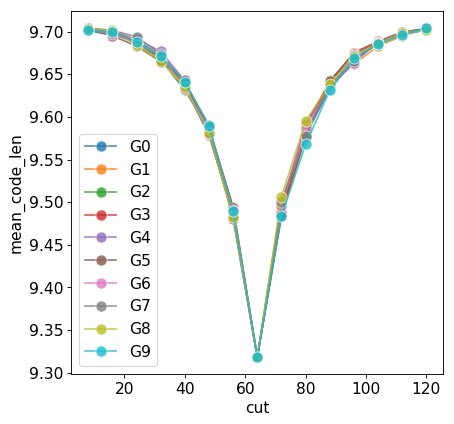}
    \includegraphics[height=0.28\textheight]{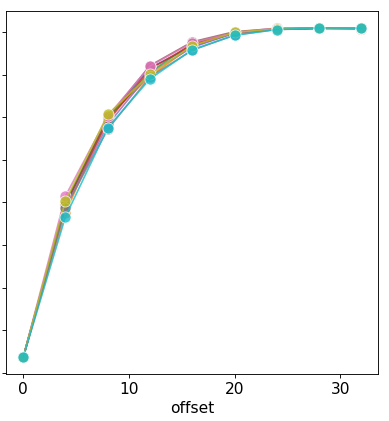}
    \caption{Three different partitions of the zachary karate club network. Sociological (upper left), minimum modularity (upper right), minimum entropy (lower)\label{zkc_part}}
    \includegraphics[width=0.4\textwidth]{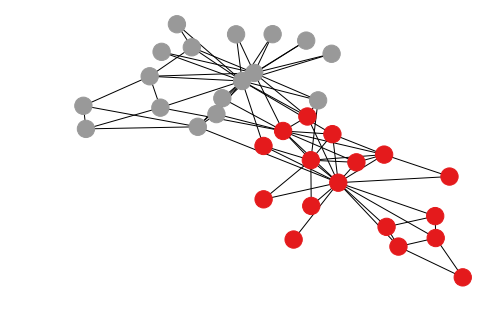}
    \includegraphics[width=0.4\textwidth]{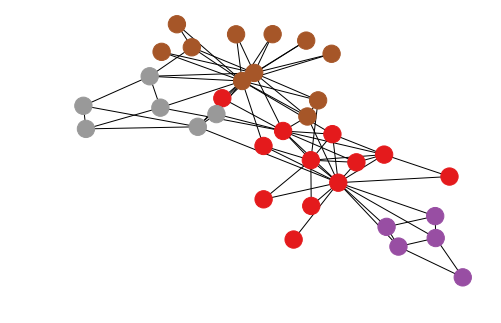}
    \includegraphics[width=0.4\textwidth]{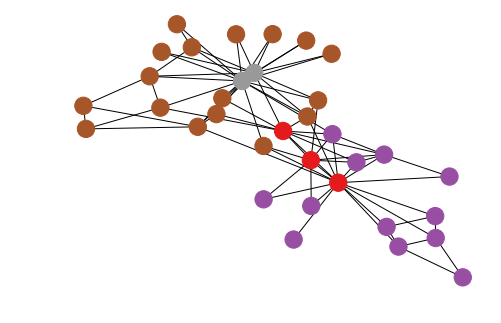}
    \caption{Mean code length for different partitions of the zachary karate club network \label{mean_code_len_zkc}}
    \includegraphics[width=\textwidth]{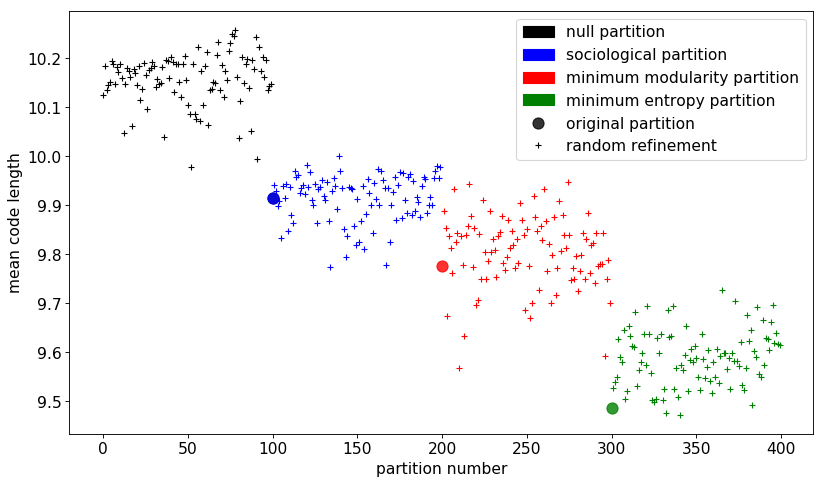}
  \end{center}
\end{figure}

Finally, we considered the performance of the mean code length when modifing blocks' sizes or shifting blocks. To do so, we generated $10$ graphs with $128$ nodes and $2800$ edges, made of two perfectly separated communities of equal size. Then, for each of these graphs, we computed the mean code length for two sequence of partitions.
\begin{itemize}
\item $S_{cut} = (B(c) = ([1, c],[c, 128]))_{c \in \{0, 8, 16, 24, \dots, 128 \}}$
\item $S_{offset} = (B(o) = ([1+o, 64+o], [1, o] \cup [65+o, 128])_{o \in \{0, 4, 8, 12, \dots, 32\}})$
\end{itemize}
Results are plotted, respectively against $c$ and $o$, on figure \ref{mean_code_len_vs_cut_offset}.

We observe that for all graphs, the minimum of mean code length is reached when $c = 64$ in the first sequence, and when $o = 0$ in the second, which both correspond to the partition $B_1$ used to generate them. This means that mean code length is robust against shifting blocks and modifying blocks' sizes.

\input{merge_split}

\section{Zachary Karate Club}

Finally, we test the mean code length quality function on the zachary karate club network. We study three different partitions of it. First of all, the sociological partition, $B_{100}$, which is the partition described in the original paper as corresponding to the sociological ground truth about communities in the karate club. $B_{200}$ is the partition obtained by minimizing the modularity using the louvain algorithm, and $B_{300}$ the partition obtained by minimizing the entropy using the graph\_tool library. Those partitions are illustrated on figure \ref{zkc_part}.

For each of these partitions, we compute the mean code length. We also do so for $100$ random partitions of the graph, with $1$ to $5$ blocks, and for each of these partitions, we compute the mean code length for $99$ random refinement of them, obtained by randomly dividing each block in two. Results are plotted on figure \ref{mean_code_len_zkc}.

We observe that the mean code length is minimum for the minimum entropy partition. All studied partitions perform better than the random ones, so the mean code length captures the fact that they reproduce part of the structure of the network. Yet, for $B_{100}$ and $B_{200}$ many of there random refinements improve the compression, sometimes by a large amount, indicating that they are not optimal. This is not the case for the minimum entropy partition $B_{300}$. There are only $2$ refinements out of $99$ which perform a little better, an issue we have seen may happen due to random fluctuations. These results are coherent with previous work showing that $B_{100}$ is actually not fully supported by statistical evidence in the network. In the case of $B_{200}$, modularity is defined based on nodes' degree, so the selected partition compensate for node degrees, which are not considered here. Finally, minimizing the entropy without correcting for the degree leads to the identification of two blocks of hubs, at the center of each sociological communities, and two blocks corresponding to their periphery. This is not necessarily what we expect, because we are used to communities defined with an implicit or explicit degree correction, but as we have not imposed constraints so far, this result corresponds to the statistical evidence present in the network.

\section{Conclusion}

In conclusion, in this paper, we have defined a new quality function, the mean code length, to evaluate node partitions. It relies on an alternative definition of the stochastic block model, as a probability distributions of edges. We make the hypothesis that the edges of the graph $G$ under study were sampled independently from the same stochastic block model probability distribution. Then, we make use of the law of great numbers and of the minimum description length principle to derive a statistical estimator of the partition used to generate $G$. The mathematical derivation of this estimator allows a clear interpretation of the partition identified. What is more, it is a basis for mathematicaly proving properties about it, for example its convergence toward the original partition.

We then test this estimator on synthetic graphs, generated with a known block structure. It shows that mean code length is able to correctly identify blocks of nodes whose internal connections are homogeneous, avoiding both the tendancy to merge distinct communities which leads to underfitting, and to split communities in smaller blocks, which leads to overfitting. Finally, we test it on different partition of the zachary karate club and the result were coherent with previous results based on statistical inference of the stochastic block model.

Those results are preliminary. This quality function should be tested more thoroughly, against graphs of various sizes and densities, with heterogeneous communities. In particular, it would be interesting to measure the density thresholds that allows stochastic block models to be recovered using this method, as at been done for other methodology.

\section*{Acknowledgments}
This work was supported by the ACADEMICS grant of the IDEXLYON, project of the Université de Lyon, PIA operated by \textbf{ANR-16-IDEX-0005}, and of the project \textbf{ANR-18-CE23-0004} (BITUNAM) of the French National Research Agency (ANR).

\bibliographystyle{unsrt}
\bibliography{edge_based_stochastic_block_model_statistical_inference}

\end{document}

%% file: merge_split.tex
\subsection{Merge / split issue}

As of today, the main stochastic block model statistical inference methodology is based on SBM considered as generative models of graphs, as explained at the beginning of section \ref{methodology presentation}. The best set of parameters $B,M$ is infered by minimizing the entropy of the set of generable graphs $\Omega_{B,M}$, as detailed in \cite{peixoto2019bayesian}. In the following, this entropy will be denoted by $\mathrm{entropy}(G, B)$, and we compute it using the python library graph tools\footnote{\url{https://graph-tool.skewed.de}}. It has been shown that this methodology can leed to a phenomenon of block inversion in graphs made of one large communities and a set of smaller ones \cite{duvivier2019minimum}. Here, we will show how the mean code length allows to overcome the issue. 

To illustrate the phenomenon on a simple example, let's consider a stochastic block model $S_1$ defined on a set of $n = 12$ nodes, partitioned in three communities: $B = ([0;5], [6;8], [9;11])$ and a probability matrix:
\[M = \begin{bmatrix}
  0.026 & 0 & 0 \\
  0 & 0.003 & 0 \\
  0 & 0 & 0.003
  \end{bmatrix}\]
We test two different partitions: the original one, $B$, and the inverse partition $B_i = ([0;2], [3;5], [6;11])$. To do so, we generate $100$ graphs $G_i$ made of $m = 378$ edges with $S_1$ and for each graph, we compute the mean code length and the entropy for both partitions. Then, for both quality function, we compute the percentage of graphs for which the original partition is identified as better than the inverse one. Results are shown in table \ref{correct_match}.

\begin{table}
  \begin{center}
    \caption{Percentage of correct match for heterogeneous graphs\label{correct_match}}
    \begin{tabular}{c|c|c}
    SBM & mean code length & entropy \\
    \hline
    $S_1$ & $96\%$ & $0\%$ \\
    $S_2$ & $100\%$ & $0\%$ 
    \end{tabular}
  \end{center}
\end{table}

While the mean code length almost always correctly identifies the original partition, the entropy of the microcanonical ensemble never does so. The graphs considered here had a very high density, which makes them not very realistic, but the same results can be obtained with low density graphs. Let's consider a stochastic block model $S_2$ on $n = 256$ nodes, partitioned in $33$ communities: one of size $128$, and $32$ of size $4$. The internal probability of the big community is $0.00006$, the one of the small communities is $0.00076$, and the probability between communities is null. As before, we generate $100$ graphs with $S_2$ and test for each of them the original partition and the inverse partition obtained by splitting the big community in $32$ small ones and merging the small ones in one big. The percentage of graph for which the mean code length (resp. the entropy) is smaller for the original partition than the inverse one is shown in table \ref{correct_match}. In this case too, the mean code length always recovers the original partition, while minimum entropy never does.